\documentclass[aps,pra,twocolumn]{revtex4-2}
\usepackage{amsmath}
\usepackage{amssymb}
\usepackage{times}
\usepackage{slashed}
\usepackage{siunitx}

\newcommand{\lbr}{\langle}
\newcommand{\rbr}{\rangle}

\allowdisplaybreaks

\begin{document}

\title{Nuclear recoil with electron vacuum polarization in muonic atoms}

	\author{Vojt\v{e}ch Patk\'o\v{s}}
	\affiliation{Faculty of Mathematics and Physics, Charles University,  Ke Karlovu 3, 121 16 Prague
		2, Czech Republic}	

\author{Mateusz Pa\'{n}tak}
\affiliation{Faculty of Physics, University of Warsaw,
	Pasteura 5, 02-093 Warsaw, Poland}
	
	\author{Krzysztof Pachucki}
	\affiliation{Faculty of Physics, University of Warsaw,
		Pasteura 5, 02-093 Warsaw, Poland}
		
\date{\today}
\begin{abstract}
Using the heavy-particle QED approach, we derive the last missing $\alpha\,(Z\alpha)^5\,m^2/M$ nuclear recoil correction in muonic atoms,
which originates from the combination of the $(Z\alpha)^5\,m^2/M$ recoil with the electron vacuum polarization. Numerical results are presented for the
$1S$, $2S$, and $2P$ states of $\mu$H, $\mu$D, $\mu^3$He$^+$, and $\mu^4$He$^+$.
For the $2P-2S$ Lamb shift in muonic helium ions, our results are of opposite sign
and about four times smaller in magnitude than the previous logarithmic estimates.
\end{abstract}
\maketitle

\section{Introduction}

Energy levels of muonic atoms are sensitive probes of nuclear structure. 
Since the Bohr radius of the muon is about 200 times smaller than that of the electron,
all nuclear structure effects are strongly enhanced. 
Thanks to this enhancement, laser spectroscopy of the $2S-2P$ transitions in $\mu$H, $\mu$D,
$\mu^3$He$^+$, and $\mu^4$He$^+$ \cite{Pohl2010,Antognini2013,Pohl2016,Krauth2021,Schuhmann2025} has yielded charge radii of light nuclei with
unprecedented precision, and X-ray spectroscopy of heavier muonic atoms aims at further
improvements for nuclei with $Z=3$--$10$ \cite{QUARTET}. 
The extraction of these nuclear charge radii from spectroscopic measurements
relies on accurate theoretical predictions of all QED and nuclear structure 
contributions \cite{Pachucki2024,Antognini2013Ann,Krauth2016,Franke2017,Diepold2018}.

In muonic atoms, the electron vacuum polarization (evp) is the dominant QED effect.
The reason is that the electron Compton wavelength is comparable to the muonic Bohr
radius, $m_e \approx m_\mu\,Z\alpha$, so that the ratio 
\begin{align}
\beta = \frac{m_e}{m_\mu\,Z\alpha}
\end{align}
is of order
one. Consequently, it is necessary to account for higher-order iterations of the evp potential
as well as for its combination with other corrections. Therefore, for muonic systems, it is
advantageous to include the evp potential directly in the nonrelativistic
Schr\"odinger equation and solve it numerically \cite{PbarSpectr}. This incorporates the evp effect
nonperturbatively into the nonrelativistic energy and avoids the systematic summation of
individual higher-order iterations. The evp has already been included nonperturbatively
in the leading relativistic $\alpha^4$ correction \cite{Veitia2004,PbarSpectr} for arbitrary masses of the constituent particles, 
as well as in the Dirac equation \cite{Michel:17} in the infinite-nuclear-mass limit and in the leading recoil correction $\sim m/M$ \cite{Rathi:26}. 
However, this all-order approach is not very efficient for light atomic systems. 

The combined effect of evp and the leading nuclear recoil at order $(Z\alpha)^5$ has so
far been known only in the logarithmic approximation \cite{JentschuraWundt2011}, and its uncertainty was
estimated at 100\% \cite{Pachucki2024}. In this work, we derive the
complete $(Z\,\alpha)^5$ recoil correction with the Uehling
potential \cite{Uehling1935} included, using the heavy-particle QED (HPQED) approach \cite{HPQED}. The Uehling potential
modifies the Bethe logarithm, the Araki--Sucher \cite{Araki1957,Sucher1958} term, and the Dirac-$\delta$ terms through the
wave function, the nonrelativistic Hamiltonian, and the energy, and gives rise to
additional effective potentials and local terms. Higher-order evp potentials
(K\"all\'en--Sabry \cite{KallenSabry1955}, Wichmann--Kroll \cite{WichmannKroll1956}) are not included
in the nuclear recoil corrections, as their contribution is
negligible; in muonic hydrogen, for instance, the K\"all\'en--Sabry correction amounts to
about 0.7\% of the Uehling correction \cite{Pachucki2024}, so we omit it here.

The paper is organized as follows. 
In Sec.~\ref{simple} we present the known $\alpha^5$ contributions for two-body systems.
In Sec.~\ref{sec:2} we present the HPQED formulas for the
leading recoil correction, and in Sec.~\ref{sec:3} we demonstrate the method by rederiving the
leading recoil Salpeter correction. In Sec.~\ref{sec:4} we extend the derivation to include the Uehling
potential. In Sec.~\ref{sec:5} we compare our result with the logarithmic approximation of
Ref.~\cite{JentschuraWundt2011}, and in Sec.~\ref{sec:6} we present numerical results for $\mu$H, $\mu$D,
$\mu^3$He$^+$, and $\mu^4$He$^+$.

\section{$\alpha^5$ correction with evp: a simple approach} \label{simple}
The QED correction of order $\alpha^5$ to the energy of a two-body system
consists of three contributions (neglecting spin-orbit and spin-spin interactions),
\begin{align}
E^{(5)} =&\ \sum_{i=1}^3 \delta_i E\,.
\end{align}
The self-energy and vacuum-polarization correction of the first particle,
\begin{widetext}
\begin{align}
\delta_1 E =&\ -\frac{2\,\alpha}{3\,\pi}\,\frac{1}{m_1^2}
\left\langle\,\vec p\,(H_0-E_0)\,\ln\bigg[\frac{2\,(H_0-E_0)}{\mu(Z\,\alpha)^2}\bigg]\,\vec{p}\right\rangle +
\frac{\alpha(Z\alpha)}{m_1^2}\,\bigg(\frac43\ln\frac{m_1\,(Z\alpha)^{-2}}{\mu} + \frac{10}{9} - \frac{4}{15}\bigg)\,\langle\delta^3(r)\rangle\,,
\end{align}
and similarly for the second particle,
\begin{align}
\delta_2 E =&\ -\frac{2\,Z^2\,\alpha}{3\,\pi}\,\frac{1}{m_2^2}
\left\langle\,\vec p\,(H_0-E_0)\,\ln\bigg[\frac{2\,(H_0-E_0)}{\mu(Z\,\alpha)^2}\bigg]\,\vec{p}\right\rangle +
\frac{Z^2\,\alpha(Z\alpha)}{m_2^2}\,\bigg(\frac43\ln\frac{m_2\,(Z\alpha)^{-2}}{\mu} + \frac{10}{9} - \frac{4}{15}\bigg)\,\langle\delta^3(r)\rangle\,.
\end{align}
Since the second particle corresponds to the nucleus, we omit the $10/9$ term because it is absorbed into the finite-nuclear-size correction.
We also drop the $-4/15$ term because it is part of the hadronic vacuum polarization.
The last part is the photon exchange correction between particles, 
which for point spin-$1/2$ particles is
\begin{align}
\delta_3 E =&\
-\frac{4\,Z\,\alpha}{3\,\pi}\,\frac{1}{m_1\,m_2}
\left\langle\,\vec p\,(H_0-E_0)\,\ln\bigg[\frac{2\,(H_0-E_0)}{\mu(Z\,\alpha)^2}\bigg]\,\vec{p}\right\rangle
-\frac{14\,(Z\,\alpha)^5}{3}\,\frac{\mu^3}{m_1\,m_2}\,\left\langle \frac{1}{4\,\pi\,(\mu\,Z\,\alpha\,r)^3}
\right\rangle \nonumber \\ &\
+\frac{(Z\,\alpha)^2}{m_1\,m_2}\,\biggl\{\frac{2}{3}\,\ln(Z\,\alpha)^{-1}+\frac{62}{9}
-\frac{2}{m_1^2-m_2^2}\,\Bigl[m_1^2\,\ln\Bigl(\frac{m_2}{\mu}\Bigr)-m_2^2\,\ln\Bigl(\frac{m_1}{\mu}\Bigr)\Bigr]
\biggr\}\,\langle\delta^3(r)\rangle\,. 
\end{align}
The nucleus is not always a spin-$1/2$ particle. 
For other values of the nuclear spin there is an extra correction \cite{Pachucki2024}, which,
however, is of order $(m_1/m_2)^3$, and we neglect it here.
Moreover, we simplify the mass dependence of the logarithmic terms in $\delta_3 E$ by omitting corrections of order $(m_1/m_2)^3$, 
\begin{align}
-\frac{2}{m_1^2-m_2^2}\,\Bigl[m_1^2\,\ln\Bigl(\frac{m_2}{\mu}\Bigr)-m_2^2\,\ln\Bigl(\frac{m_1}{\mu}\Bigr)\Bigr]
 \approx&\ -2\,\ln\Bigl(\frac{m_1}{\mu}\Bigr)\,. 
\end{align}
Thus, the sum of all terms after simplifications is
\begin{align}
E^{(5)} =&\ \label{6}
	-\frac{2\,\alpha}{3\,\pi}\,\biggl(\frac{1}{m_1}+\frac{Z}{m_2}\biggr)^2
\left\langle\,\vec p\,(H_0-E_0)\,\ln\bigg[\frac{2\,(H_0-E_0)}{\mu(Z\,\alpha)^2}\bigg]\,\vec{p}\right\rangle 	
\nonumber \\ &\
+\frac{\alpha(Z\alpha)}{m_1^2}\,\bigg(\frac43\ln\frac{m_1\,(Z\alpha)^{-2}}{\mu} + \frac{10}{9} - \frac{4}{15}\bigg)\,\langle\delta^3(r)\rangle
+\frac{Z^2\,\alpha(Z\alpha)}{m_2^2}\,\frac43\ln\frac{m_2\,(Z\alpha)^{-2}}{\mu}\,\langle\delta^3(r)\rangle
\nonumber \\ &\ 
-\frac{14\,(Z\,\alpha)^5}{3}\,\frac{\mu^3}{m_1\,m_2}\,\left\langle \frac{1}{4\,\pi\,(\mu\,Z\,\alpha\,r)^3}\right\rangle	
+\frac{(Z\,\alpha)^2}{m_1\,m_2}\,\biggl[\frac{2}{3}\,\ln(Z\,\alpha)^{-1}+\frac{62}{9}
-2\,\ln\Bigl(\frac{m_1}{\mu}\Bigr)  \biggr]\,\langle\delta^3(r)\rangle\,.
\end{align}
Let us now include evp in this formula. 
This is achieved by including the leading evp in the nonrelativistic Hamiltonian $H_0$, the energy $E_0$,
and the wave function $\phi$.
The potential $V$ is the sum of the Coulomb potential and the Uehling potential.
Eq.~(\ref{6}) is thus modified to
\begin{align}
E^{(5)}_\mathrm{evp} =&\ \label{7}
	-\frac{2\,\alpha}{3\,\pi}\,\biggl(\frac{1}{m_1}+\frac{Z}{m_2}\biggr)^2
\left\langle\,\vec p\,(H_0-E_0)\,\ln\bigg[\frac{2\,(H_0-E_0)}{\mu(Z\,\alpha)^2}\bigg]\,\vec{p}\right\rangle_\mathrm{evp} 	
\nonumber \\ &\
+\frac{\alpha}{m_1^2}\,\bigg(\frac43\ln\frac{m_1\,(Z\alpha)^{-2}}{\mu} + \frac{10}{9} - \frac{4}{15}\bigg)\,\frac{\langle\nabla^2 V\rangle_\mathrm{evp}}{4\,\pi}
-\frac{4\alpha}{15m_1^2}\,\frac{\langle\nabla^2 V_\mathrm{VP}\rangle_\mathrm{evp}}{4\,\pi}
+\frac{Z^2\,\alpha}{m_2^2}\,\frac43\ln\frac{m_2\,(Z\alpha)^{-2}}{\mu}\,\frac{\langle\nabla^2 V\rangle_\mathrm{evp}}{4\,\pi}
\nonumber \\ &\ 
-\frac{14\,(Z\,\alpha)^5}{3}\,\frac{\mu^3}{m_1\,m_2}\,\left\langle \frac{1}{4\,\pi\,(\mu\,Z\,\alpha\,r)^3}\right\rangle_\mathrm{evp}	
+\frac{Z\,\alpha}{m_1\,m_2}\,\biggl[\frac{2}{3}\,\ln(Z\,\alpha)^{-1}+\frac{62}{9}
-2\,\ln\Bigl(\frac{m_1}{\mu}\Bigr)  \biggr]\,\frac{\langle\nabla^2 V\rangle_\mathrm{evp}}{4\,\pi}
\,,
\end{align}
\end{widetext}
where the notation $\langle A\rangle_\mathrm{evp}$ stands for the expectation value of the 
operator $A$, calculated using the wave function obtained by solving the Schr\"odinger
equation with evp contained in the Hamiltonian. 
We will suppress writing this subscript explicitly in the following sections.
The extra term proportional to $4\alpha/(15 m_1^2)$ on the second
line comes from the combined muon and electron vacuum-polarization contribution
and accounts for the two possible orderings of the VP insertions on the same exchanged photon.
The simple replacement $V_C\rightarrow V_C+V_\mathrm{VP}$, which we performed
when going from Eq.~(\ref{6}) to Eq.~(\ref{7}), does not account for the complete
evp contribution.
The additional evp correction beyond the above formula is the topic of this work. To derive it, we use the HPQED approach,
which is valid to first order in $m_1/m_2$, and write $m_1=m_\mu=m$ for the muon mass and $m_2=M$ for the nuclear mass.

\section{Leading recoil correction $E_\mathrm{rec}$}\label{sec:2}
The exact-in-$Z\alpha$ HPQED formula for the leading recoil correction in hydrogenic atoms in the so-called $\mathcal{F}_3$ form \cite{HPQED} is
\begin{align}
E_\mathrm{rec} =&\
\frac{i}{M} \int_s \frac{d\omega}{2\,\pi}\,  \langle\phi| D_T^j(\omega) \,G(E_D + \omega)\,D_T^j(\omega) |\phi\rangle\,,
\end{align}
where the Feynman contour is assumed and the subscript $s$ denotes the symmetric integration around the pole at $\omega= 0$.
Further, 
\begin{align}
G(E) = [E-H_D]^{-1}
\end{align}
is the Dirac-Coulomb Green function with
\begin{align}
H_D = \vec\alpha\cdot\vec p + \beta m - \frac{Z\,\alpha}{r}\,, \label{02}
\end{align}
where $\vec \alpha$, $\beta$ are Dirac matrices,
and
\begin{align}
D_T^j(\omega,\vec r) =&\ -4\pi Z\alpha \, \alpha^i \, G_{T}^{ij}(\omega,\vec{r})\,, \label{13}
\end{align}
with $G_T$ being the photon propagator in the temporal gauge,
\begin{align}
G_T^{ij}(\omega, \vec k) =&\ \frac{1}{k^2}\,\biggl(\delta^{ij}-\frac{k^i\,k^j}{\omega^2}\biggr)\,.
\end{align}

For the derivation of particular terms, it may be convenient to use a different (but equivalent) form of the recoil correction 
$E_\mathrm{rec}$ as the starting point. Namely, $E_\mathrm{rec}$ in the $\mathcal{F}_2$ form is
\begin{align}
E_\mathrm{rec} =&\
-\frac{i}{M} \int_s \frac{d\omega}{2\,\pi}\,  \langle\phi| \big[p^j(V_C) -\omega\,D_C^j(\omega)\big] \,G(E_D + \omega)\,
\nonumber \\ &\ \times
\big[ p^j(V_C) + \omega\, D_C^j(\omega)\big] |\phi\rangle\,\frac{1}{\omega^2}\,,
\end{align}
and
in the $\mathcal{F}_1$ form it is
\begin{align}
E_\mathrm{rec} =&\
\frac{i}{M} \int \frac{d\omega}{2\,\pi}\,  \langle\phi| \big[ p^j-D_C^j(\omega)\big] \,G(E_D + \omega)
\nonumber \\ \times &\
\big[p^j-D_C^j(\omega)\big] |\phi\rangle\,.
\end{align}
In the actual calculation, we expand the exact formula for $E_\mathrm{rec}$ in powers of
$Z\alpha$ to obtain the correction of order $(Z \alpha)^5$. First, we show that the HPQED approach
can be used to recover the well-known Salpeter correction $E_\mathrm{rec}^{(5)}$ \cite{Salpeter1952,Eides2001}. 
Then, we extend the derivation to include the Uehling potential nonperturbatively in the  
$E_\mathrm{rec}^{(5)}$ contribution.

\section{Salpeter correction $E^{(5)}_\mathrm{rec}$}\label{sec:3}
In this section we derive the Salpeter correction $E_\mathrm{rec}^{(5)}$ (without evp).
In order to calculate it, we use dimensional regularization, in which the momentum integrals take the form
\begin{align}
\int \frac{d^3 k}{(2\,\pi)^3} f(k)  \rightarrow &\ m^{3-d}\int \frac{d^d k}{(2\,\pi)^d}f(k) \nonumber \\
= &\ \frac{\Omega_d}{(2\,\pi)^d}\,m^{3-d}\int dk\,k^{d-1} f(k)\,,
\end{align}
with the solid angle 
\begin{align}
\Omega_d =&\ \frac{2\,\pi^{d/2}}{\Gamma(d/2)}\,.
\end{align}
Next, we split $E^{(5)}_\mathrm{rec}$ into three parts:
\begin{align}
E^{(5)}_\mathrm{rec} = E_H + E_M + E_L\,.
\end{align}
The high-energy part $E_H$ comes from the region where the photon momentum is of order $\omega \sim m$, the middle-energy part from the region $\omega \sim \alpha\, m$, and
the low-energy part from the region of photon momenta $\omega \sim \alpha^2\,m$. Each part is calculated in $d$ dimensions with 
$d=3-2\,\epsilon$ and expanded in small $\epsilon$. 
The singular $1/\epsilon$ terms cancel in the sum, leading to a finite result at $d=3$.

\subsection{Low-energy part $E_L$}
The low-energy contribution comes from the momenta $\omega\sim \alpha^2\, m$
of the photon exchanged between the nucleus and muon. Specifically, using the $\mathcal{F}_1$ form, 
it comes from the single transverse photon exchange 
\begin{widetext}
\begin{align}
E_L =&\
 \frac{i}{M} \int \frac{d\omega}{2\,\pi} \, \, \lbr \phi | \big[ p^j - D_C^j(\omega)\big] \,
G(E_D+\omega) \, \big[ p^j - D_C^j(\omega)\big] | \phi \rbr \nonumber \\
%= &\ -\frac{i}{M} \int \frac{d\omega}{2\,\pi} \, \lbr \phi |  p^j  \,
%G(E_D+\omega) \, D_C^j(\omega) + D_C^j(\omega) \, G(E_D+\omega) \, p^j | \phi \rbr \nonumber \\
%= &\ \frac{i}{M}\,(4\pi Z\alpha) \int \frac{d\omega}{2\,\pi} \, \lbr \phi |  p^j  \,
%G(E_D+\omega) \,\alpha^k G_C^{kj}(\omega,\vec r) + \alpha^k G_C^{kj}(\omega,\vec r) \, G(E_D+\omega) \, p^j | \phi \rbr \nonumber \\
= &\ \frac{i}{M}\,(4\pi Z\alpha)\,m^{3-d}\int \frac{d\omega}{2\,\pi} \int\frac{d^d k}{(2\pi)^d} 
\frac{1}{\omega^2 - \vec k^2}\bigg(\delta^{kj} - \frac{k^k k^j}{\vec k^2}\bigg)
 \, \lbr \phi |  p^j  \,
G(E_D+\omega) \,\alpha^k e^{i\vec k\cdot\vec r} + \alpha^k e^{i\vec k\cdot\vec r} \, G(E_D+\omega) \, p^j | \phi \rbr \nonumber \\
= &\ \frac{1}{M} (4\pi Z\alpha)\,m^{3-d} \int\frac{d^d k}{(2\pi)^d} \frac{1}{2k}\bigg(\delta^{kj} - \frac{k^k k^j}{\vec k^2}\bigg)
\,  \, \lbr \phi |  p^j  \,
G(E_D-k) \,\alpha^k e^{i\vec k\cdot\vec r} + \alpha^k e^{i\vec k\cdot\vec r} \, G(E_D-k) \, p^j | \phi \rbr\,.
\end{align}
\end{widetext}
For the correction of order $\alpha^5$ and $k\sim \alpha^2\,m$,
we may use the nonrelativistic limit of the fermion propagator, 
\begin{align}
G(E_0-k) = (E_0 - H_0 - k)^{-1},
\end{align}
the dipole approximation $e^{i\vec k\cdot\vec r}\approx 1$, and the nonrelativistic approximation $\alpha^k\rightarrow p^k/m$, so that
%\begin{widetext}
\begin{align}
E_L= &\ 
\frac{(4\pi Z\alpha)}{m\,M} m^{3-d} \int\frac{d^d k}{(2\pi)^d} \frac{1}{k}\bigg(\delta^{kj} - \frac{k^k k^j}{\vec k^2}\bigg)\, 
\nonumber \\ &\ \times
\lbr \phi |  p^j  \,\frac{1}{E_0-H_0-k} \,p^k  | \phi \rbr
%\nonumber \\= &\ 
%\frac{(4\pi Z\alpha)}{m\,M} \frac{(d-1)}{d} m^{3-d} \int\frac{d^d k}{(2\pi)^d} \frac{1}{k}\, \lbr \phi |  p^j  \,\frac{1}{E_0-H_0-k} \,p^j  | \phi \rbr
%\nonumber \\= &\ 
%\frac{(4\pi Z\alpha)}{m\,M} \frac{(d-1)}{d} m^{3-d}\,\frac{\Omega_d}{(2\,\pi)^d} \int d k\, k^{d-2}\, \lbr \phi |  p^j  \,\frac{1}{E_0-H_0-k} \,p^j  | \phi \rbr
%\nonumber \\= &\ 
%\frac{(4\pi Z\alpha)}{m\,M} \frac{(d-1)}{d} m^{3-d}\,\frac{\Omega_d}{(2\,\pi)^d}\,\frac{\pi}{\sin[(d-2)\,\pi]}\,  \lbr \phi |  p^j  \, (H_0-E_0)^{d-2} \,p^j  | \phi \rbr
%\nonumber \\= &\ 
%\frac{(4\pi Z\alpha)}{M} 
%\bigg\{\frac{(d-1)}{d}\,\frac{1}{(2\,\pi)^d}\,\frac{2\,\pi^{d/2}}{\Gamma(d/2)}\,\frac{\pi}{\sin[(d-2)\,\pi]}\bigg\}
%\, \lbr \phi |  p^j  \, \bigg(\frac{H_0-E_0}{m}\bigg)^{d-2} \,p^j  | \phi \rbr
\nonumber \\= &\ 
\frac{(4\pi Z\alpha)}{M}\,(4\,\pi)^\epsilon\,\Gamma(1+\epsilon)\, \frac{1}{6\,\pi^2}\,\bigg(\frac{5}{3} + \frac{1}{\epsilon} - 2\,\ln 2\bigg) 
\nonumber \\ &\ \times
\,\lbr \phi |  p^j  \, \bigg(\frac{H_0-E_0}{m}\bigg)^{1-2\,\epsilon} \,p^j  | \phi \rbr\,.
\end{align}
We consistently drop the factor $(4\,\pi)^\epsilon\,\Gamma(1+\epsilon)$ in all parts; therefore we also drop it here and obtain
\begin{align}
	E_L = &\  
	-\frac{4 \,Z\alpha}{3\,\pi\,m\,M}\lbr\phi| p^i\,(H_0 - E_0)\ln\bigg[\frac{2(H_0-E_0)}{m(Z\alpha)^2}\bigg]\,p^i|\phi\rbr
	\nonumber \\ &\
	 + \frac{(Z\alpha)^2}{m\,M}
	\bigg(\frac{20}{9} + \frac{4}{3\,\epsilon}
	+\frac{8}{3}\ln(Z\alpha)^{-2} \bigg)\,\lbr\phi|\delta^d(r)|\phi\rbr\,.
\end{align}

\subsection{Middle-energy part $E_M$}
In this region, the photon momentum is of order $\omega \sim \alpha\, m$. 
Therefore, using the $\mathcal{F}_3$ form, one may
approximate $G(E_D + \omega)$ by the free fermion propagator and obtain
%\begin{widetext}
\begin{align}
E_M =&\
\frac{i}{M} \int_s \frac{d\omega}{2\,\pi}\,  \langle\phi| D_T^j(\omega) \,G(E_D + \omega)\,D_T^j(\omega) |\phi\rangle \nonumber \\
=  &\ \frac{i\,(4\pi Z\alpha)^2}{2\,M\,m} \int_s \frac{d\omega}{2\,\pi}\,
\langle\phi| \,G_T^{ij}(\omega,\vec r) \,\alpha^i\,\alpha^k \,G^{kj}_T(\omega,\vec r) |\phi\rangle\,.
\end{align}
%\end{widetext}
Using
\begin{equation}
\alpha^i\,\alpha^k = \sigma^i\,\sigma^k = \delta^{ik} + i \epsilon^{ikl}\,\sigma^l,
\end{equation}
and the nonrelativistic approximation for the wave function $\phi$ we obtain
\begin{align}
E_M =&\
 -\frac{(4\pi Z\alpha)^2}{2\,M\,m} (m^2)^{3-d}\int_s \frac{d\omega}{2\,\pi\,i}\,\int\frac{d^d k_1}{(2\pi)^d}\int\frac{d^dk_2}{(2\pi)^d} 
 \nonumber \\ &\ \times
 \frac{1}{\omega^2-\vec k_1^2}\,\biggl(\delta^{ij}-\frac{k_1^i\,k_1^j}{\omega^2}\biggr)\frac{1}{\omega^2-\vec k_2^2}\,\biggl(\delta^{ij}-\frac{k_2^i\,k_2^j}{\omega^2}\biggr) 
 \nonumber \\ &\ \times
\langle\phi| e^{i (\vec k_1+\vec k_2)\cdot\vec r} |\phi\rangle\,.
\end{align}
Integrations over $\omega$ and $k_1$ lead to (after shifting the momentum $\vec k_2 \rightarrow
\vec q-\vec k_1$)

\begin{align}
&\ \int_s \frac{d\omega}{2\,\pi\,i}\,\int\frac{d^d k_1}{(2\pi)^d}\int\frac{d^dk_2}{(2\pi)^d}
 \frac{1}{\omega^2-\vec k_1^2}\,\biggl(\delta^{ij}-\frac{k_1^i\,k_1^j}{\omega^2}\biggr)
\nonumber \\ &\ \times
\frac{1}{\omega^2-\vec k_2^2}\,\biggl(\delta^{ij}-\frac{k_2^i\,k_2^j}{\omega^2}\biggr)\,e^{i(\vec k_1+\vec k_2)\cdot\vec r}\nonumber \\
 &\ =  \frac{1}{\pi^2}\int\frac{d^dq}{(2\pi)^d}\,\bigg(-\frac{1}{24} + \frac{7}{24\,\epsilon} - \frac{7}{12}\,\ln q\bigg) \,e^{i\vec q\cdot\vec r}\,.
\end{align}
Inserting this into the equation for $E_M$, we obtain
\begin{align}
E_M =&\ 
  \frac{(Z\,\alpha)^2}{M\,m}\bigg[
 \int\frac{d^3q}{(2\pi)^3}\,\bigg(\frac{1}{3} + \frac{14}{3}\,\ln \frac{q}{m}\bigg)
 \langle\phi| e^{i\,\vec q\cdot\vec r}|\phi\rangle 
 \nonumber \\ &\
 - \frac{7}{3\,\epsilon}\,\langle\phi|\delta^d(r)|\phi\rangle\bigg]\,,
 \end{align}
 where we set $d\rightarrow3$ in the first line of the last equality,
 since this part is regular.
 Bearing in mind that (au stands for atomic units) 
 \begin{align}
 &\ \int\frac{d^3q}{(2\pi)^3}\,\ln \frac{q}{m}\, \langle e^{i\,\vec q\cdot\vec r}\rangle =
  -\frac{1}{4\,\pi}\bigg\langle \frac{1}{r^3}\bigg\rangle + \langle \delta^3(r)\rangle(1-\ln m)
  \nonumber \\ &\ =
   -\frac{(m\,Z\,\alpha)^3}{4\,\pi}\bigg\langle \frac{1}{r^3}\bigg\rangle_\mathrm{au} 
   + [1+\ln(Z\,\alpha)]\,\langle \delta^3(r)\rangle\,,
  \end{align} 
  where the momentum integration is performed as described in Appendix \ref{appendix:B}.
  Then $E_M \equiv \langle H_M\rangle$ becomes
  \begin{align}
  E_M  =&\ \frac{(Z\,\alpha)^2}{M\,m}\bigg[\bigg(5+\frac{14}{3}\,\ln(Z\,\alpha)\bigg)\,\langle\delta^3(r)\rangle
  \nonumber\\ &\
  -(m\,Z\,\alpha)^3\,\frac{7}{6\,\pi}\,\bigg\langle \frac{1}{r^3}\bigg\rangle_\mathrm{au} 
  - \frac{7}{3\,\epsilon}\,\langle\phi|\delta^d(r)|\phi\rangle\bigg]\,.
 \end{align}
 This equation serves as a definition of the middle-energy effective operator
 $H_M$.

\subsection{High-energy part $E_H$}

The high-energy part comes from the region of photon momenta $\omega\sim m$,
and the corresponding contribution is proportional to the wave function at the origin:
\begin{widetext}
\begin{align}
E_H =&\ \frac{i}{M}\,(Z\alpha)^2\,\phi^2(0)\,m^{3-d}\int\frac{d\omega}{2\,\pi}\int\frac{d^dk}{(2\pi)^d}\frac{(4\pi)^2}{k^4}\bigg(\delta^{ik}-\frac{k^i k^k}{\omega^2}\bigg)
\bigg(\delta^{jk}-\frac{k^j k^k}{\omega^2}\bigg)\mathrm{Tr}\bigg[\gamma^i\frac{1}{\slashed{k}+\slashed{t}-m}\,\gamma^j\frac{(\gamma^0+I)}{4} \bigg] \nonumber \\
%= &\ \frac{i}{M}\,(Z\alpha)^2\,\phi^2(0)\,m^{3-d}\int\frac{d\omega}{2\,\pi}\int\frac{d^dk}{(2\pi)^d}\frac{(4\pi)^2}{k^4}\bigg(\delta^{ij}-\frac{k^i k^j}{\omega^2}\bigg)
%\bigg(\delta^{ij}-\frac{k^i k^j}{\omega^2}\bigg)\frac{\omega}{k^2 + 2\,m\,\omega} \nonumber \\
= &\ \frac{i}{M}\,(Z\alpha)^2\,\phi^2(0)\,m^{3-d}\int\frac{d\omega}{2\,\pi}\int\frac{d^dk}{(2\pi)^d}\frac{(4\pi)^2}{k^4}\bigg(d - 2\frac{\vec k^2}{\omega^2}+\frac{\vec k^4}{\omega^4}\bigg)\frac{\omega}{k^2 + 2\,m\,\omega} \nonumber \\
= &\ \frac{(Z\,\alpha)^2}{m\,M}\langle\phi|\delta^d(r)|\phi\rangle
\bigg(\frac{1}{\epsilon} - \frac13\bigg)
\equiv \langle H_H\rangle\,.
\end{align}
where $t=(m,\vec 0)$ (on-shell muon at rest).
\end{widetext}

\subsection{Complete $E^{(5)}_\mathrm{rec}$}
The complete recoil contribution $E^{(5)}_\mathrm{rec}$ is the sum of the low-energy, middle-energy, and
high-energy parts. All singular contributions cancel and the final result is finite.
In atomic units, it is equal to
\begin{align}
E^{(5)}_\mathrm{rec} =&\ E_L+E_M+E_H
%\nonumber \\ =&\
%-\frac{4 \,Z\alpha}{3\,\pi\,m\,M}\lbr\phi| p^i\,(H_0 - E_0)\ln\bigg[\frac{2(H_0-E_0)}{m(Z\alpha)^2}\bigg]\,p^i|\phi\rbr
%	 + \frac{(Z\alpha)^2}{\pi\,m\,M}
%	\bigg(\frac{20}{9} + \frac{4}{3\,\epsilon}
%	+\frac{8}{3}\ln(Z\alpha)^{-2} \bigg)\,\lbr\phi|\delta^d(r)|\phi\rbr
%\nonumber \\ &\
%+  \frac{(Z\,\alpha)^2}{M\,m}\bigg[\bigg(5+\frac{14}{3}\,\ln(Z\,\alpha)\bigg)\,\langle\delta^3(r)\rangle
%  -(m\,Z\,\alpha)^3\,\frac{7}{6\,\pi}\,\bigg\langle \frac{1}{r^3}\bigg\rangle_\mathrm{au} 
%  - \frac{7}{3\,\epsilon}\,\delta^d(r)\bigg]
%+  \frac{(Z\,\alpha)^2}{m\,M}\langle\phi|\delta^d(r)|\phi\rangle\,\bigg(\frac{1}{\epsilon} - \frac13\bigg)	
\nonumber \\ =&\
\frac{m^2}{M}\,\frac{(Z\,\alpha)^5}{\pi}\bigg[
\bigg(\frac{62}{9}+\frac{1}{3}\ln(Z\alpha)^{-2} \bigg)\,\pi\,\lbr \delta^3(r)\rbr_\mathrm{au}
\nonumber \\ &\
-\frac{8}{3\,n^3}\,\ln k_0(n,l)
- \frac{7}{6}\,\bigg\langle\frac{1}{r^3}\bigg\rangle_\mathrm{au}\bigg]\,,
\end{align}
where
$\ln k_0(n,l)$
is the Bethe logarithm defined in Eq.~(\ref{def:BetheLogarithm}).
We have thus recovered the leading recoil Salpeter correction. 

\section{$E^{(5)}_\mathrm{rec}$ with vacuum polarization}\label{sec:4}
Next, we include the one-loop vacuum polarization in the interaction.
This is achieved by modifying the photon propagator as

\begin{align}
	G_T^{ij}(\omega, \vec k) =&\ \frac{1-\bar{\omega}(k^2)}{k^2}\,\biggl(\delta^{ij}-\frac{k^i\,k^j}{\omega^2}\biggr)\,, 
\end{align}
where
\begin{align}	
\bar{\omega}(k^2)	=&\ \frac{\alpha}{\pi}\,k^2\int_2^\infty \frac{d\xi}{\xi}\,
	\frac{1}{(m_e^2\,\xi^2-k^2)}
	\frac23\sqrt{1-\frac{4}{\xi^2}}\bigg(1+\frac{2}{\xi^2}\bigg)\,.
\end{align}
The Dirac Hamiltonian $H_D$, with the corresponding relativistic energy $E_D$, now explicitly
contains the Uehling potential,
\begin{align}
	H_D =&\ \vec\alpha\cdot\vec p + \beta m + V,
\end{align}
where
\begin{align}
V =&\  V_C + V_\mathrm{VP}\,,\\
V_C =&\  -\frac{Z\,\alpha}{r}\,,\\
V_\mathrm{VP} =&\ -\frac{\alpha\,Z\alpha}{\pi}\int\frac{d^dk}{(2\pi)^d}\int_2^\infty \frac{d\xi}{\xi}
\,\frac{4\pi}{(\vec{ k}^2+m_e^2\,\xi^2)}\nonumber \\ &\
\times\frac{2}{3}\sqrt{1-\frac{4}{\xi^2}}\bigg(1+\frac{2}{\xi^2}\bigg)\, e^{i\vec k\cdot\vec r}\,.
\end{align}
Similarly, the modified nonrelativistic Hamiltonian $H_0$, with the nonrelativistic energy $E_0$, is
\begin{align}
	H_0 	=\frac{p^2}{2\mu} + V\,.
\end{align}
At our level of precision, we set $\mu=m$ in the following, since our results are valid to first order in $m/M$.
We restore the $\mu$ dependence in the final result for the part which can be
identified as a direct generalization of the Salpeter correction.

As in the previous section, we separate the calculation
into the low-energy, middle-energy, and high-energy parts, which come from
photon momenta of order $\alpha^2\,m$, $\alpha\,m$, and $m$, respectively.
The low-energy contribution $E_L$ has the form 
%\begin{widetext}
\begin{align}
	E_L
	= &\ \frac{(4\pi Z\alpha)}{m\,M} \frac{(d-1)}{d} m^{3-d}
\nonumber\\ &\ \times	
\,  \int\frac{d^d k}{(2\pi)^d} \frac{1}{k}\, \lbr \phi |  p^j  \,\frac{1}{(E_0-H_0-k)} \,p^j  | \phi \rbr \nonumber \\
%	= &\ \frac{(4\pi Z\alpha)}{m\,M} m^{3-d}\frac{(d-1)}{d}\,\frac{\pi^{1-\frac{d}{2}}\,\csc(d\,\pi)}{2^{d-1}\,\Gamma\big(\frac{d}{2}\big)}\,\lbr\phi| p^j\,\big(H_0-E_0\big)^{d-2}\,p^j|\phi\rbr\nonumber \\
	= &\ -\frac{(4\pi Z\alpha)}{m\,M} \frac{\Gamma(d)\,\Gamma(2-d)}{(4\pi)^\frac{d}{2}\,\Gamma\big(\frac{d}{2}+1\big)}\,m^{3-d}
	\nonumber \\ &\ \times
	\lbr\phi| p^j\,\big(H_0-E_0\big)^{d-2}\,p^j|\phi\rbr\,.
\end{align}

%\end{widetext}
Expanding this for small $\epsilon$, we get 
%and transforming into reduced atomic units by
%\begin{align}
%\vec r \rightarrow &\ (m\, Z\alpha)^{-\frac{1}{4-d}}\,\vec r \nonumber \\
%\vec k \rightarrow &\ (m\, Z\alpha)^{\frac{1}{4-d}}\,\vec k  \nonumber \\
%m_e \rightarrow &\ ( m\, Z\alpha)^{\frac{1}{4-d}}\,\beta
%\end{align}
%with $\beta$ being parameter of the order of one, we get
\begin{align}
E_L =&\ E_{L1} + E_{L2}\,,\\
E_{L1} = &\  -\frac{4 \,Z\alpha}{3\,\pi\,m\,M}\lbr\phi| p^i\,(H_0 - E_0)\ln\bigg[\frac{2(H_0-E_0)}{m(Z\alpha)^2}\bigg]\,p^i|\phi\rbr \nonumber \\
	%-\frac{8\,(Z\alpha)^5\,m^2}{3\,\pi\,M\,n^3}\ln k_0
+ &\ \frac{Z\alpha}{\pi\,m\,M} \bigg(\frac{5}{9} + \frac{1}{3\epsilon}
+\frac{2}{3}\ln(Z\alpha)^{-2} \bigg)\,
\lbr\phi|\nabla^2 V_C|\phi\rbr\,, \\	
E_{L2} =&\ 
%\frac{Z\alpha}{\pi\,m\,M} \bigg(\frac{5}{9} + \frac{1}{3\epsilon}
	   %      +\frac{2}{3}\ln(Z\alpha)^{-2} \bigg)\,\lbr\phi|\nabla^2 V|\phi\rbr 
%\nonumber \\ =&\
\frac{Z\alpha}{\pi\,m\,M} \bigg(\frac{5}{9} + \frac{1}{3\epsilon}
	         +\frac{2}{3}\ln(Z\alpha)^{-2} \bigg)\, \lbr\phi|\nabla^2
	          V_\mathrm{VP}|\phi\rbr  \,.	         
\end{align}
Let us recall that the nonrelativistic Hamiltonian $H_0$, the energy $E_0$, and the
wave function $\phi$ now implicitly contain the Uehling potential.

%Here $V_\mathrm{a.u.}=V_{C,\mathrm{a.u.}}+V_\mathrm{VP,a.u.}$ where
%\begin{align}
%V_{C,\mathrm{a.u.}} = &\ -\frac{1}{r}\nonumber \\
%V_\mathrm{VP,a.u.} = &\ \frac{\alpha}{\pi}\int\frac{d^dk}{(2\pi)^d}\int_2^\infty d\xi\,\frac{4\pi}{(\vec{ k}^2+\beta^2\,\xi^2)}\,\frac{2}{3\,\xi}\sqrt{1-\frac{4}{\xi^2}}\bigg(1+\frac{2}{\xi^2}\bigg)\, e^{i\vec k\cdot\vec r}
%\end{align}

For the middle-energy part, we have
\begin{align}
E_M =&\ E_{M1}+E_{M2}\,,\\
E_{M1} =&\ \langle\phi| 
%V_\mathrm{VP}\,\frac{1}{(E_0-H_0)'} 
H_M |\phi\rangle\,,
\end{align}
and
\begin{widetext}
\begin{align}
E_{M2} =&\
	-\frac{(4\pi Z\alpha)^2}{2\,M\,m} (m^2)^{3-d}\int_s \frac{d\omega}{2\,\pi\,i}\,\int\frac{d^d k_1}{(2\pi)^d}\int\frac{d^dk_2}{(2\pi)^d}
	\frac{1}{\omega^2-\vec k_1^2}\,\biggl(\delta^{ij}-\frac{k_1^i\,k_1^j}{\omega^2}\biggr)\frac{1}{\omega^2-\vec k_2^2}\,\biggl(\delta^{ij}-\frac{k_2^i\,k_2^j}{\omega^2}\biggr)\nonumber \\
	&\ \times \big[-\bar{\omega}(k_1^2)-\bar{\omega}(k_2^2)\big]
	\langle\phi| e^{i (\vec k_1+\vec k_2)\cdot\vec r} |\phi\rangle
\nonumber \\ =&\
	-\frac{(4\pi Z\alpha)^2}{M\,m} (m^2)^{3-d}\int_s \frac{d\omega}{2\,\pi\,i}\,\int\frac{d^d k_1}{(2\pi)^d}\int\frac{d^dk_2}{(2\pi)^d}
	\frac{1}{\omega^2-\vec k_1^2 -m_e^2\,\xi^2}\,\biggl(\delta^{ij}-\frac{k_1^i\,k_1^j}{\omega^2}\biggr)\frac{1}{\omega^2-\vec k_2^2}\,\biggl(\delta^{ij}-\frac{k_2^i\,k_2^j}{\omega^2}\biggr)\nonumber \\
	&\times \frac{\alpha}{\pi}\int_2^\infty \frac{d\xi}{\xi}\,\frac23\sqrt{1-\frac{4}{\xi^2}}\bigg(1+\frac{2}{\xi^2}\bigg)
	\langle\phi| e^{i (\vec k_1+\vec k_2)\cdot\vec r} |\phi\rangle\,.	
\end{align}
\end{widetext}
To evaluate $E_{M2}$, we use the integration formula
\begin{align}
&\	\int\frac{d\omega}{2\pi}\int\frac{d^dk}{(2\pi)^d}\frac{\omega^\gamma}{\big[(k-q)^2+\omega^2\big]^a\,\big[k^2+\omega^2+\xi^2\big]^b} \nonumber \\
	= &\ 2^{-d-1}\,\pi^{-1-\frac{d}{2}}\,\frac{\Gamma\big(\frac{\gamma+1}{2}\big)\Gamma
		(\mu)\Gamma(b-\mu)}{\Gamma(b)\Gamma\big(\frac{\gamma+d+1}{2}\big)}\xi^{-2\mu}
		\nonumber \\ &\ \times
		\,{}_2F_1\bigg(a,\mu;a+b-\mu;-\frac{q^2}{\xi^2}\bigg)\,,
\end{align}
where $\gamma$ is even, and $\mu = a + b - (\gamma + d +1)/2$.
Next, we split $E_{M2}$ into two parts by introducing a cutoff $\Lambda$ in the $\xi$ integration,
\begin{equation}
	E_{M2} = E_{M2}^A + E_{M2}^B\,.
\end{equation}
The first part corresponds to small $\xi<\Lambda$. Here, we expand in small $\epsilon$ before performing the $\xi$ integration and obtain
\begin{widetext}
\begin{align}
	E_{M2}^A = &\ \frac{\alpha}{\pi}\frac{(Z\alpha)^2}{m\,M}m^{3-d}\int\frac{d^dk}{(2\pi)^d}
	\int_2^\Lambda d\xi \,
	\frac{2}{3\,\xi}\sqrt{1-\frac{4}{\xi^2}}\bigg(1+\frac{2}{\xi^2}\bigg)
	\bigg[-\frac23+\frac{4k^2}{3(k^2+m_e^2\xi^2)} + \frac{1}{\epsilon}\bigg(-2
	-\frac{4k^2}{3\,(k^2+m_e^2\xi^2)}\bigg) \nonumber \\
&\	+\bigg(2 + \frac{2m_e^2 \xi^2}{k^2}
	+\frac{8k^2}{3(k^2+m_e^2\xi^2)}\bigg)\ln\frac{k^2+m_e^2\,\xi^2}{m_e^2\,\xi^2} + \bigg(2 + \frac{4k^2}{3(k^2+m_e^2\xi^2)}\bigg) \ln\frac{m_e^2\xi^2}{m^2}\bigg]\,\langle\phi|e^{i\,\vec k\cdot\vec r}|\phi\rangle	\,.
\end{align}
\end{widetext}
The second part, $E_{M2}^B$, comes from the region of large $\xi>\Lambda$. 
In this integration region, we can set the argument of the hypergeometric function ${}_2F_1\big(a,\mu;a+b-\mu;-q^2/\xi^2\big)$ to zero, $q^2/\xi^2\rightarrow0$. 
We can also use the approximation
 \begin{align}
 \frac{2}{3\xi}\sqrt{1-\frac{4}{\xi^2}}\big(1+\frac{2}{\xi^2}\big)\approx\frac{2}{3\xi}.
 \end{align}
 After combining the two parts $E_{M2}^A$ and $E_{M2}^B$, the dependence on the cutoff $\Lambda$ cancels exactly and we can set $\Lambda\rightarrow\infty$. The final result for
the middle-energy part $E_{M2}$ is
\begin{widetext}
\begin{align}
	E_{M2} = &\ \frac{\alpha}{\pi}\frac{(Z\alpha)^2}{m\,M}\int\frac{d^3k}{(2\pi)^3}
	\int_2^\infty d\xi \,
	\frac{2}{3\,\xi}\sqrt{1-\frac{4}{\xi^2}}\bigg(1+\frac{2}{\xi^2}\bigg) \bigg[
	\bigg(2 
	+\frac{8k^2}{3\big(k^2+m_e^2\xi^2\big)}\bigg)\ln\bigg(\frac{  k^2}{m_e^2\xi^2}+1 \bigg) 
	\nonumber \\
	&\ 
	+\frac{4k^2}{3\big(k^2+m_e^2\xi^2\big)} \ln \frac{m_e^2 \xi^2}{m^2}
	+2\,
	\frac{m_e^2 \xi^2}{k^2}\bigg(\ln\bigg(\frac{  k^2}{m_e^2\xi^2}+1 \bigg) -
	\frac{  k^2}{m_e^2\xi^2} 
	\bigg)\bigg]\,\langle\phi|e^{i\,\vec k\cdot\vec r}|\phi\rangle
	\nonumber \\ &\
	+\frac{\alpha}{\pi}\frac{Z\alpha}{m\,M}
\bigg[ -\frac{76}{27} + \frac{\pi^2}{9} 
%\nonumber \\ &\ 
- \frac{28}{9}\ln\frac{m_e}{m}
-\frac43\ln^2\frac{m_e}{m} -\frac{2}{3\,\epsilon^2} + \frac{1}{\epsilon}\bigg(\frac{14}{9}+\frac43\ln\frac{m_e}{m}\bigg)\bigg] \frac{\langle\phi|\nabla^2 V_{C}|\phi\rangle}{4\pi} \nonumber \\
&\ + \frac{Z\alpha}{\pi\,m\,M}\bigg(\frac13 - \frac{1}{3\epsilon}\bigg)\langle\phi|\nabla^2 V_{VP}|\phi\rangle\,. 
\end{align}
\end{widetext}
Finally, the high-energy part is
\begin{align}
E_H =&\ E_{H1} + E_{H2}\,,\\
E_{H1} = &\ \langle\phi|
%V_\mathrm{VP}\,\frac{1}{(E_0-H_0)'}\,
H_H|\phi\rangle\,,\\
E_{H2} = &\ \frac{i}{M}\,(4\pi\,Z\alpha)\int\frac{d\omega}{2\,\pi}\int\frac{d^dk}{(2\pi)^d}\frac{\big(-2\bar{\omega}(k^2)\big)}{k^4}
\nonumber \\ &\ \times
\bigg(\delta^{ij}-\frac{k^i k^j}{\omega^2}\bigg)
	\bigg(\delta^{ij}-\frac{k^i k^j}{\omega^2}\bigg)\frac{\omega}{k^2 + 2\,m\,\omega}
	\nonumber \\ &\ \times
	\,\langle\phi|\nabla^2 V_C |\phi\rangle\,. 
	%\nonumber \\
	% + &\ \frac{i}{M}\,(4\pi\,Z\alpha)\,\langle\phi| \nabla^2 V_{VP}|\phi\rangle\int\frac{d\omega}{2\,\pi}\int\frac{d^dk}{(2\pi)^d}\frac{1}{k^4}\bigg(\delta^{ij}-\frac{k^i k^j}{\omega^2}\bigg)
	%\bigg(\delta^{ij}-\frac{k^i k^j}{\omega^2}\bigg)\frac{\omega}{k^2 + 2\,m\,\omega}
\end{align}
Here, we can use the asymptotic form of $\bar{\omega}(k^2)$ valid for large $k^2$:
\begin{equation}\label{46}
	\bar{\omega}(k^2) \approx \frac{\alpha}{3\pi}\bigg(\frac53 - \ln\bigg(-\frac{k^2}{m_e^2}\bigg) \bigg)\,.
\end{equation}
This part can then be evaluated to
\begin{align}\label{EH2}
E_{H2} = &\ \frac{\alpha}{\pi}\frac{Z\alpha}{m\,M}\frac{\langle\phi|\nabla^2 V_{C}|\phi\rangle}{4 \pi}
\bigg(-\frac{40}{27}-\frac{\pi^2}{3}+\frac49\ln\frac{m_e}{m} 
\nonumber \\ &\
+\frac{2}{3\,\epsilon^2}
-\frac{14}{9\,\epsilon} - \frac{4}{3\,\epsilon}\ln\frac{m_e}{m}\bigg)\,.
\end{align}

We can now combine the low-energy, middle-energy, and high-energy contributions.
The singular terms containing $1/\epsilon$ and $1/\epsilon^2$ cancel and the final result
is finite:
\begin{widetext}
\begin{align}\label{48}
E^{(5)}_\mathrm{evp,rec} =&\  E_L + E_M + E_H
\nonumber \\ =&\
\delta_\mathrm{VP} E^{(5)}_\mathrm{rec}
+ \frac{Z\alpha}{\pi\,m\,M}\bigg(\frac89
	+\frac{2}{3}\ln(Z\alpha)^{-2} + \frac23\ln\frac{m_e}{m}\bigg)\langle\phi|\nabla^2 V_{\mathrm{VP}}|\phi\rangle
\nonumber \\ &\
+\frac{\alpha}{\pi}\frac{Z\alpha }{m\,M}\bigg\{Z\alpha\int\frac{d^3k}{(2\pi)^3}
	\int_2^\infty d\xi \,
	\frac{2}{3\,\xi}\sqrt{1-\frac{4}{\xi^2}}\bigg(1+\frac{2}{\xi^2}\bigg) \bigg[
	\bigg(2 
	+\frac{8k^2}{3\big(k^2+m_e^2\xi^2\big)}\bigg)\ln\bigg(\frac{  k^2}{m_e^2\xi^2}+1 \bigg) 
	\nonumber \\
	&\ 
	+\frac{4k^2}{3\big(k^2+m_e^2\xi^2\big)} \ln \xi^2
	+2\,
	\frac{m_e^2 \xi^2}{k^2}\bigg(\ln\bigg(\frac{  k^2}{m_e^2\xi^2}+1 \bigg) -
	\frac{  k^2}{m_e^2\xi^2} 
	\bigg)\bigg]\,\langle\phi|e^{i\,\vec k\cdot\vec r}|\phi\rangle
	\nonumber \\ &\
	+	\frac{1}{6\,\pi}\,
	\bigg(-\frac{58}{9} - \frac{\pi^2}{3} - 4\ln \frac{m_e}{m}
	- 2\,\ln^2\frac{m_e}{m} \bigg)\,\langle\phi|\nabla^2 V_{\mathrm{C}}|\phi\rangle
	\bigg\}\,,
\end{align}
\end{widetext}
where $\delta_\mathrm{VP} E^{(5)}_\mathrm{rec}$ is the leading recoil Salpeter contribution  
with the Uehling potential $V_\mathrm{VP}$ included in the nonrelativistic potential $V$, the wave function $\phi$, and $H_0-E_0$ in the Bethe logarithm.
The remaining terms in Eq.~(\ref{48}) represent the extra vacuum-polarization contributions
to the $\alpha^5$ recoil correction.
We now transform $E^{(5)}_\mathrm{evp,rec} $ to atomic units by
\begin{align}
\vec r \rightarrow &\ (m\, Z\alpha)^{-1}\vec r\,, \nonumber \\
\vec k \rightarrow &\ (m\, Z\alpha)\,\vec k\,,  \nonumber \\
m_e \rightarrow &\ ( m\, Z\alpha)\,\beta\,,
\end{align}
with $\beta$ being a parameter of order one. We define effective potentials (in atomic units)
\begin{widetext}
\begin{align}
	V_C =&\ -\frac{1}{r}\,,\\
	V_\mathrm{VP} =&\
	-\frac{\alpha}{\pi}\int_2^\infty d\xi\,\frac{2}{3\,\xi}\sqrt{1-\frac{4}{\xi^2}}\bigg(1+\frac{2}{\xi^2}\bigg)
	\int\frac{d^3k}{(2\pi)^3}e^{i\vec k\cdot\vec r}\,\frac{4\pi}{\vec{ k}^2+\beta^2\,\xi^2}
	\nonumber \\ =&\ 
	-\frac{\alpha}{\pi}\int_2^\infty d\xi\,\frac{2}{3\,\xi}\sqrt{1-\frac{4}{\xi^2}}\bigg(1+\frac{2}{\xi^2}\bigg)
	\frac{\exp(-\beta\,\xi\,r)}{r}\,, \\
	V_\mathrm{VP1} =&\ \frac{\alpha}{\pi}\int_2^\infty d\xi\,\frac{2}{3\,\xi}\sqrt{1-\frac{4}{\xi^2}}\bigg(1+\frac{2}{\xi^2}\bigg)
	\int\frac{d^3k}{(2\pi)^3}e^{i\vec k\cdot\vec r}\,\frac{4\pi}{\vec{ k}^2+\beta^2\,\xi^2}\,\ln(\beta^2\,\xi^2)
	\nonumber \\ =&\
	\frac{\alpha}{\pi}\int_2^\infty d\xi\,\frac{2}{3\,\xi}\sqrt{1-\frac{4}{\xi^2}}\bigg(1+\frac{2}{\xi^2}\bigg)\,\ln(\beta^2\,\xi^2)\,
	\frac{\exp(-\beta\,\xi\,r)}{r}\,, \\
	V_\mathrm{VP2} =&\ \frac{\alpha}{\pi}\int_2^\infty d\xi\,\frac{2}{3\,\xi}\sqrt{1-\frac{4}{\xi^2}}\bigg(1+\frac{2}{\xi^2}\bigg)
	\int\frac{d^3k}{(2\pi)^3}e^{i\vec k\cdot\vec r}\,\frac{4\pi}{k^2}\,\ln\bigg(\frac{  k^2}{\beta^2\xi^2}+1 \bigg)
	\nonumber \\ =&\
	 -\frac{\alpha}{\pi}\int_2^\infty d\xi\,\frac{2}{3\,\xi}\sqrt{1-\frac{4}{\xi^2}}\bigg(1+\frac{2}{\xi^2}\bigg)
	\,\frac{2}{r}\,\mathrm{Ei}(-\beta\,\xi\,r)\,,	
	\\ 
	V_\mathrm{VP3} =&\ \frac{\alpha}{\pi}\int_2^\infty d\xi\,\frac{2}{3\,\xi}\sqrt{1-\frac{4}{\xi^2}}\bigg(1+\frac{2}{\xi^2}\bigg)
	\int\frac{d^3k}{(2\pi)^3}e^{i\vec k\cdot\vec r}\;
	\frac{4\pi\,\beta^2 \xi^2}{k^4}\bigg[\ln\bigg(\frac{  k^2}{\beta^2\xi^2}+1 \bigg) - \frac{  k^2}{\beta^2\xi^2} \bigg]
	\nonumber \\ =&\
	\frac{\alpha}{\pi}\int_2^\infty d\xi\,\frac{2}{3\,\xi}\sqrt{1-\frac{4}{\xi^2}}\bigg(1+\frac{2}{\xi^2}\bigg)
	\beta^2\,\xi^2\,r\,\bigg( \mathrm{Ei}(-\beta\,\xi\,r) + \frac{e^{-\beta\,\xi\,r}}{\beta\,\xi\,r} -\frac{e^{-\beta\,\xi\,r}}{\beta^2\,\xi^2\,r^2}\bigg)\,,
	\\
	V_\mathrm{VP4} =&\ \frac{\alpha}{\pi}\int_2^\infty d\xi\,\frac{2}{3\,\xi}\sqrt{1-\frac{4}{\xi^2}}\bigg(1+\frac{2}{\xi^2}\bigg)
	\int\frac{d^3k}{(2\pi)^3}e^{i\vec k\cdot\vec r}\,\frac{4\pi}{k^2+\beta^2\,\xi^2}
	\,\ln\bigg(\frac{  k^2}{\beta^2\xi^2}+1 \bigg)	 
\nonumber \\ =&\	
-\frac{\alpha}{\pi}\int_2^\infty d\xi\,\frac{2}{3\,\xi}\sqrt{1-\frac{4}{\xi^2}}\bigg(1+\frac{2}{\xi^2}\bigg)\,	
	\frac{e^{-\beta\,\xi\,r}}{r}\,\big[\gamma + e^{2\,\beta\,\xi\,r}\,\mathrm{Ei}(-2\,\beta\,\xi\,r) + \ln(\beta\,\xi\,r/2) \big]\,,
\end{align}
\end{widetext}
where the details of their derivation are given in Appendix \ref{appendix:A}.
The result for $E^{(5)}_\mathrm{evp,rec}$ in atomic units can then be written
in the compact form (omitting the index au for operators in atomic units)
\begin{align}
E^{(5)}_\mathrm{evp,rec} = &\ 
\frac{m^2}{M}\,\frac{(Z\,\alpha)^5}{\pi}\bigg\{
-\frac{7}{6} \bigg\lbr \frac{1}{r^3} \bigg\rbr +\bigg(\frac{62}{9}-\frac{2}{3}\ln(Z\alpha) \bigg)
\nonumber\\ &\ \times
\langle\pi\,\delta^{3}(r)\rangle-\frac{8}{3\,n^3}\,\ln k_0(n,l)
-\langle \nabla^2V_\mathrm{SVP}\rangle 
\nonumber \\ &\ 	
+ \bigg(\frac89 -\frac{2}{3}\ln(Z\alpha)\bigg)\langle\nabla^2 V_{\mathrm{VP}}\rangle
- \frac{\alpha}{\pi}\,\frac{4}{3} \bigg(\frac{29}{9} + \frac{\pi^2}{6} 
\nonumber \\ &\
+ 2\,\ln \big[\beta\,Z\alpha\big]
+ \ln^2\big[\beta\,Z\alpha\big] \bigg)\,\langle \pi\,\delta^{3}(r)\rangle\bigg\}\,,
\end{align}
where $\ln k_0(n,l)$ is the Bethe logarithm with the vacuum
polarization contained in the wave function, the nonrelativistic Hamiltonian $H_0$, and the nonrelativistic energy $E_0$, 
and
\begin{align}
V_\mathrm{SVP} =&\ 
\frac{1}{3}\,V_\mathrm{VP1}  + \frac{1}{2}\,V_\mathrm{VP2}  + \frac{1}{2}\,V_\mathrm{VP3} + \frac{2}{3}\,V_\mathrm{VP4}
\nonumber \\ =&\
\frac{\alpha}{\pi}\int_2^\infty d\xi\,\frac{2}{3\,\xi}\sqrt{1-\frac{4}{\xi^2}}\bigg(1+\frac{2}{\xi^2}\bigg)\,\frac{e^{-\beta\,\xi\,r}}{r}
\nonumber \\ &\
\times
\bigg[ \frac{1}{3}\,\ln(\beta^2\,\xi^2) -
 e^{\beta\,\xi\,r}\,\mathrm{Ei}(-\beta\,\xi\,r)	
\nonumber \\ &\
-\frac{1}{2}\,\big(1 - \beta\,\xi\,r -\beta^2\,\xi^2\,r^2\,e^{\beta\,\xi\,r}\,\mathrm{Ei}(-\beta\,\xi\,r)  \big)
\nonumber \\ &\
-\frac{2}{3}\,\big(e^{2\,\beta\,\xi\,r}\,\mathrm{Ei}(-2\,\beta\,\xi\,r) + \ln(\beta\,\xi\,r/2)+\gamma \big)\bigg]
\end{align}
is an additional effective vacuum-polarization potential.
Let us now again split this result into two parts: $\delta_\mathrm{VP} E^{(5)}_\mathrm{rec}$,
which is the Salpeter formula $E^{(5)}_\mathrm{rec}$ with the Uehling potential included in the wave function,
the nonrelativistic Hamiltonian $H_0$, and the energy $E_0$; and an extra contribution $\delta E^{(5)}$,
which contains all the additional terms:
\begin{align}
E^{(5)}_\mathrm{evp,rec} = &\ 	\delta_\mathrm{VP} E^{(5)}_\mathrm{rec} + \delta E^{(5)}\,, \label{evprec} \\
\delta_\mathrm{VP} E^{(5)}_\mathrm{rec} = &\ \frac{\mu^3}{m\,M}\frac{(Z\,\alpha)^5}{\pi}\bigg\{
 -\frac{7}{6} \bigg\lbr \frac{1}{r^3} \bigg\rbr +\bigg(\frac{62}{9}-\frac{2}{3}\ln(Z\alpha)
 \nonumber \\ &\ -2\ln\frac{m}{\mu} \bigg)
 \,\frac{\langle\nabla^2 V\rangle}{4}-\frac{8}{3\,n^3}\,\ln k_0(n,l)\bigg\} \,,\label{deltaVPE5} \\
\delta E^{(5)} = &\ \frac{m^2}{M}\,\frac{(Z\,\alpha)^5}{\pi}\bigg\{
-\langle \nabla^2V_\mathrm{SVP}\rangle 	
+ \bigg(-\frac56 -\frac{1}{2}\ln(Z\alpha)
\nonumber \\ &\ +\frac12\ln\frac{m}{\mu}\bigg)
\langle\nabla^2 V_{\mathrm{VP}}\rangle  - \frac{\alpha}{\pi}\,\frac{4}{3} \bigg(\frac{29}{9} + \frac{\pi^2}{6}
\nonumber \\ &\  + 2\,\ln \big[\beta\,Z\alpha\big]
+ \ln^2\big[\beta\,Z\alpha\big] \bigg)\,\langle \pi\,\delta^{3}(r)\rangle\bigg\}\,,\label{58}
\end{align}
where in the part $\delta_\mathrm{VP} E^{(5)}_\mathrm{rec}$ we have restored the same dependence on the masses of both particles, including the logarithmic term, as in the traditional result for the Salpeter correction of order $\alpha^5$ in Eq.~(\ref{7}).
The Bethe logarithm is defined as
\begin{align}
	\label{def:BetheLogarithm}
	&\ln[k_{0}(n, l)] \equiv \frac{n^3}{2 \mu^3 (Z \alpha)^4}
	\nonumber\\& 
	\times\left\langle \phi\left| \vec{p} \, (H_{0}-E_{0}) \ln\biggl[\frac{2(H_{0}-E_{0})}{\mu (Z\alpha)^2} \biggr] \, \vec{p}\, \right|\phi \right\rangle\,.
\end{align}
For the contribution of the extra effective operators $\delta E^{(5)}$, we cannot rigorously show that the full mass dependence 
is given only by the factor $\mu^3/(m M)$
without additional logarithmic terms, since our starting HPQED formulas are valid
only to first order in $m/M$. 
For the $\delta E^{(5)}$ part, we therefore use only the leading recoil approximation and omit the higher-order recoil corrections.
These corrections are expected to be of relative order $m/M$ with respect to $\delta E^{(5)}$, i.e., below $0.03\,\mu$eV for the $2P-2S$ transitions and below $0.3\,\mu$eV for all levels considered, and are neglected.
Because Eq.~(\ref{deltaVPE5}) contains $\langle\nabla^2 V\rangle $ instead of $\langle\nabla^2 V_C\rangle $, 
the corresponding $\langle\nabla^2 V_\mathrm{VP}\rangle $ term, including the logarithm added when restoring the mass dependence, is subtracted in Eq.~(\ref{58}).

The complete contribution of order $\alpha^5$ with the Uehling potential included is then
given by the sum of the contribution $E^{(5)}_\mathrm{evp}$ of Eq.~(\ref{7}),
part of which is contained in the expression $\delta_\mathrm{VP} E^{(5)}_\mathrm{rec}$
of Eq.~(\ref{deltaVPE5}), and the extra first-order recoil contribution $\delta E^{(5)}$
of Eq.~(\ref{58}):
\begin{equation}
	E_\mathrm{evp,tot}^{(5)} = E^{(5)}_\mathrm{evp} + \delta E^{(5)}\,.\label{61}
\end{equation}

\section{Comparison to former work}\label{sec:5}

We can compare our result 
for the combined nuclear recoil and evp contribution
with the earlier derivation 
by Jentschura and Wundt \cite{JentschuraWundt2011},
which was obtained in the logarithmic approximation.
The corresponding contributions are given by Eqs.~(35), (38), and (43) of \cite{JentschuraWundt2011},
\begin{align}
E_L = &\
-\frac{m^2}{M}\,\frac{(Z\,\alpha)^5}{\pi}\,\frac{2}{3}\ln(Z\alpha)\,\langle \nabla^2(V_C+V_\mathrm{VP})\rangle\,,
\\
E_S=&\ 
\frac{m^2}{M}\,\frac{(Z\,\alpha)^5}{\pi}\,\frac{\alpha}{\pi}\,\frac{4}{3}\,\langle \pi\,\delta^{3}(r)\rangle\,
\big[-\ln^2(4\,Z\,\alpha\,\beta^2) 
\nonumber\\ &\
+ \ln(2\,\beta)\,\ln(4\,Z\,\alpha\,\beta^2)\big]
\nonumber \\ \approx&\
-\frac{m^2}{M}\,\frac{(Z\,\alpha)^5}{\pi}\,\frac{\alpha}{\pi}\,\frac{4}{3}\,\langle \pi\,\delta^{3}(r)\rangle\,
\big[\ln^2(Z\,\alpha\,\beta)
\nonumber \\ &\
 + \ln(Z\,\alpha\,\beta)\,\ln(8\,\beta)\big]\,,
\\
E_W=&\
\frac{m^2}{M}\,\frac{(Z\,\alpha)^5}{\pi}\,2\,\ln(Z\,\alpha)\,\langle \pi\,\delta^{3}(r)\rangle\,.
\end{align}
This can be compared with the logarithmic part of our results,
\begin{align}
E^{(5),\mathrm{log}}_\mathrm{evp,rec} = &\ 
\frac{m^2}{M}\,\frac{(Z\,\alpha)^5}{\pi}\bigg\{
\bigg(2-\frac{8}{3} \bigg)\,\ln(Z\alpha)\,\langle\pi\,\delta^{3}(r)\rangle
\nonumber\\ &\	
- \frac{2}{3}\ln(Z\alpha)\langle\phi|\nabla^2 V_{\mathrm{VP}}|\phi\rangle
\nonumber \\ &\ 
- \frac{\alpha}{\pi}\,\frac{4}{3} \bigg( 2\ln \big[\beta\,Z\alpha\big]
+ \ln^2\big[\beta\,Z\alpha\big] \bigg)\,\langle \pi\,\delta^{3}(r)\rangle\bigg\}\,.
\end{align}
The first two lines correspond to the sum $E_L+E_W$. 
However, the third line of our result disagrees with $E_S$. Namely,
we agree on the quadratic term, but the coefficient of the single-logarithmic term differs from
that of Ref.~\cite{JentschuraWundt2011}. 
In that paper, the high-energy contribution [their Fig.~4(e),(f)] was not evaluated, under the assumption that it only cancels the overlap parameter
(see the discussion in the first paragraph of Sec.~4.3 of Ref.~\cite{JentschuraWundt2011}). 
With the vacuum-polarization insertion, however, the high-energy region depends on $\ln(m/m_e)$ [see Eq.~(\ref{EH2})] and contributes to the single-logarithmic term, 
which therefore cannot be obtained from the seagull part alone.

\section{Numerical results}\label{sec:6}

We evaluated $E^{(5)}_\mathrm{evp,tot}$ for the $1S$, $2S$, and $2P$ states of muonic
hydrogen, muonic deuterium, and the muonic helium ions $\mu^3$He$^+$ and $\mu^4$He$^+$,
treating the nucleus as point-like. Fundamental constants and particle masses were taken
from CODATA 2022 \cite{CODATA2022}. The corresponding values of the parameter $\beta = m_e/(m_\mu\,Z\alpha)$
are $\beta = 0.6628$ for $\mu$H and $\mu$D, and $\beta = 0.3314$ for the muonic helium
ions. 

The nonrelativistic wave functions with the Uehling potential included nonperturbatively
were obtained with the PbarSpectr program \cite{PbarSpectr}. The Bethe logarithms with nonperturbative
inclusion of the Uehling potential were calculated with the PbarBetheLog program; the
details of this calculation will be presented in a separate paper. The differences between
the Bethe logarithms with and without the Uehling potential are listed in
Table~\ref{bethelog}. For $P$ states this difference is very small, since the $2P$ wave
function vanishes at the origin and is only weakly affected by the short-range Uehling
potential. The numerical uncertainties of the Bethe logarithms and of all other matrix
elements are smaller than the digits presented in the results below.

Table~\ref{E5recVP} presents the $\alpha^5$ correction to the energy with and without the Uehling
potential, together with its breakdown into the individual contributions of
Eq.~(\ref{61}). As a check, the results obtained without the Uehling
potential reproduce the known $\alpha^5$ QED values \cite{Pachucki2024,Eides2001}
up to the neglected $(m/M)^3$ terms. For all systems and states considered, the evp correction 
due to the extra effective operators, $\delta E^{(5)}$, is at most approximately
$2\,\mu$eV. The dominant contribution to $\Delta$ comes from the
combined evp and muon self-energy correction.

Table~\ref{LambShift} shows the $\alpha^5$ recoil contributions to the $2P-2S$ Lamb shift
with and without the Uehling potential, and their difference, which represents the
combined effect of evp and nuclear recoil. Since the evp corrections for $2P$ states are
small, the difference is dominated by the $2S$ state. For muonic hydrogen and deuterium,
our results, $0.18\,\mu$eV and $0.07\,\mu$eV, are consistent with the previous
logarithmic estimate \cite{JentschuraWundt2011}.
For the muonic helium ions, we obtain $-1.26\,\mu$eV for $\mu^3$He$^+$ and
$-1.01\,\mu$eV for $\mu^4$He$^+$. These values are of opposite sign and about four
times smaller in magnitude than the estimates of Ref.~\cite{JentschuraWundt2011}, from which they differ by
$1.3\,\sigma$ of the uncertainty assigned in Ref.~\cite{Pachucki2024}.

The reason for this disagreement is that the nonlogarithmic terms are of the same size as
the logarithmic ones. For muonic atoms,
$\ln(m/m_e)\approx 5.3$ is not a large parameter. Already in the local term of
Eq.~(\ref{58}), the constant $29/9+\pi^2/6\approx 4.9$ amounts to almost 30\% of the
logarithmic part $2\ln(m_e/m)+\ln^2(m_e/m)\approx 17.8$, and the contributions of the
effective potentials $V_\mathrm{VP}$ and $V_\mathrm{SVP}$ are of comparable size. The
logarithmic approximation is therefore not reliable for the combined evp and recoil effects in
muonic atoms.

For the determination of the nuclear charge radius $r_C$ from the Lamb shift of muonic atoms \cite{Pachucki2024},
the difference between our result and the logarithmic approximation from Ref.~\cite{JentschuraWundt2011}
changes the result for $r_C$ by less than 2\% of its uncertainty for all systems and is
thus negligible. Nevertheless, its importance lies in removing the second-largest uncertainty (after
hadronic vacuum polarization) of the point-nucleus QED theory.

Our results also yield the evp corrections to the muon self-energy and muon vacuum polarization. 
They are contained in Table~\ref{E5recVP} as part of $E^{(5)}_{\mathrm{evp}}$
and are not presented separately.
Our results for
the combined muonic and electronic vacuum polarization are in agreement
with the results of Sec.~III.H of Ref.~\cite{Pachucki2024}.
Regarding the combined correction of the muon self-energy and electronic vacuum polarization,
the results presented in Ref.~\cite{JentschuraWundt2011}, which were used for the
determination of the nuclear charge radii in Ref.~\cite{Pachucki2024}, contained a mistake.
This mistake was corrected in the recent work \cite{Ohayon:24}, and our numerical results
are consistent with theirs once the spin-orbit part, not included in Eq.~(\ref{7}), is removed
from their results. 
The relative difference is less than $0.3\%$ for all systems and may arise from the higher-order Uehling contributions included in our nonperturbative treatment.
Note that the result for $\mu^3$He${}^+$ in Eq.~(32c) of Ref.~\cite{Ohayon:24} contains a typographical
error: as follows from the coefficients in its Tables I--III, it should read $-0.067631$~meV
instead of $-0.067312$~meV, i.e., a digit appears to have been dropped.
Our result for this system agrees with the corrected value to within $0.3\%$ after subtraction of the spin-orbit contribution.

\begin{widetext}

	\begin{table}[!ht]
	\centering
	\caption{Difference between the Bethe logarithms with and without the Uehling potential for various states and systems, $\Delta \ln k_0 = \ln k_0(C+U) - \ln k_0 (C)$.
	The Bethe logarithm is defined in Eq.~(\ref{def:BetheLogarithm}) and is calculated
	with the full reduced-mass dependence.}
	\label{bethelog}
	\renewcommand{\arraystretch}{1.35}
	
	\begin{tabular}{
			c
			|S[table-format=-0.11]
			|S[table-format=-0.11]
			|S[table-format=-0.11]
			|S[table-format=-0.11]
		}
		\hline\hline
		State
		& {$\mu$H}
		& {$\mu$D}
		& {$\mu^3$He${^+}$}
		& {$\mu^4$He${^+}$}
		\\
		
		\hline
		
		${1S}$
		& 0.02892264(5)
		& 0.02978166(5)
		& 0.04277274(5)
		& 0.04294896(5)
		\\
		${2S}$
		& 0.02631609(5)
		& 0.02700134(5)
		& 0.03701179(5)
		& 0.03714571(5)
		\\
		
		${2P}$
		& -0.00003390(2)
		& -0.00003597(2)
		& -0.00007534(2)
		& -0.00007600(2)
		\\
		
		\hline\hline
	\end{tabular}
	
\end{table}

%\end{widetext}

\begin{table}[!ht]
	\centering
	\caption{Complete contribution of order $\alpha^5$ with evp included, in meV. Here,
		$E^{(5)}$ is the $\alpha^5$ contribution without evp given by Eq.~(\ref{6}), 
		$E^{(5)}_{\mathrm{evp}}$ is the contribution of Eq.~(\ref{7}) obtained by including
		the Uehling potential in the nonrelativistic wave function, Hamiltonian $H_0$, and energy $E_0$. Further,
		$\delta E^{(5)}$ is the contribution of the extra operators in Eq.~(\ref{58}),
		$E^{(5)}_{\mathrm{evp,tot}} = E^{(5)}_{\mathrm{evp}} + \delta E^{(5)}$ is the complete $\alpha^5$ contribution with the Uehling potential included, and 
		$\Delta 
		= E^{(5)}_{\mathrm{evp,tot}} - E^{(5)}$ is the evp correction.	}
	\label{E5recVP}
	\renewcommand{\arraystretch}{1.35}
	
	\begin{tabular}{
			c
			c
			|S[table-format=-0.6]
			|S[table-format=-0.6]
			|S[table-format=-0.6]
			|S[table-format=-0.6]
			|S[table-format=-0.6]
			|S[table-format=-0.6]
		}
		\hline\hline
		System
		& State
		& {$E^{(5)}$  }
		& {$E^{(5)}_{\mathrm{evp}}-E^{(5)}$  }
		& {$\delta E^{(5)}$  }
		& {$E^{(5)}_{\mathrm{evp,tot}}$  }	
		& {$\Delta $  }

		\\
		
		\hline
		
		$\mu$H
		& ${1S}$
		& 5.493880
		& 0.024713
		&-0.002136 
		& 5.516458
		& 0.022578
		\\
		& ${2S}$
		& 0.706591
		& 0.002661
		&-0.000257
		& 0.708995
		& 0.002405
		\\
		& ${2P}$
		& 0.000363
		&-0.000036
		&-0.000005
		& 0.000322
		&-0.000041
		\\

		\hline
		
		$\mu$D
		& ${1S}$
		& 6.141416
		& 0.029175
		&-0.001012
		& 6.169579
		& 0.028162
		\\
		
		& ${2S}$
		& 0.787706
		& 0.003129
		&-0.000121
		& 0.790715
		& 0.003008
		\\
		& ${2P}$
		& 0.001679
		&-0.000042
		&-0.000003
		& 0.001634
		&-0.000045
		\\
		
		\hline
		
		$\mu^3$He${^+}$
		& ${1S}$
		&85.948053
		& 0.634645
		& 0.001231
		&86.583929
		& 0.635876
		\\
		
		& ${2S}$
		&11.099419
		& 0.063856
		& 0.000350
		&11.163624
		& 0.064205
		\\
		
		& ${2P}$
		& 0.021193
		&-0.001068
		&-0.000133
		& 0.019992
		&-0.001201
		\\
		
		\hline
		
		$\mu^4$He${^+}$
		& ${1S}$
		&86.968368
		& 0.649831
		& 0.001148
		&87.619347
		& 0.650979
		\\
		
		& ${2S}$
		&11.218743
		& 0.065411
		& 0.000292
		&11.284446
		& 0.065703
		\\
		
		& ${2P}$
		& 0.028980
		&-0.001083
		&-0.000101
		& 0.027796
		&-0.001184
		\\
		
		\hline\hline
	\end{tabular}
\end{table}

%\begin{widetext}

	\begin{table}[!ht]
	\centering
	\caption{Comparison of the $\alpha^5$ recoil contributions to the $2P-2S$ Lamb shift in different systems with and without the nonperturbative inclusion of the Uehling potential, and their difference, in meV (computed from unrounded values).
	Here, $E^{(5)}_{\mathrm{rec}}$ is the Salpeter correction, 
	and $E^{(5)}_{\mathrm{evp,rec}}$ is the corresponding contribution with the Uehling potential.}
	\label{LambShift}
	\renewcommand{\arraystretch}{1.35}
	
	\begin{tabular}{
			c
			|S[table-format=-0.8]
			|S[table-format=-0.8]
			|S[table-format=-0.8]
			|S[table-format=-0.8]
		}
		\hline\hline
		
		& {$\mu$H}
		& {$\mu$D}
		& {$\mu^3$He${^+}$}
		& {$\mu^4$He${^+}$}
		\\
		
		\hline
		
		$E^{(5)}_{\mathrm{evp,rec}}(2P-2S)$  
		& -0.04439
		& -0.02641
		& -0.55850
		& -0.43363
		\\
		$E^{(5)}_{\mathrm{rec}}(2P-2S)$  
		& -0.04457
		& -0.02649
		& -0.55724
		& -0.43262
		\\ \hline
		
		Difference
		& 0.00018
		& 0.00007
		&-0.00126
		&-0.00101
		
		\\
		
		Refs.~\cite{JentschuraWundt2011,Pachucki2024}
		& 0.00014(14)
		& 0.00009(9)
		& 0.0049(49)
		& 0.0039(39)
		\\
		
		\hline\hline

	\end{tabular}
	
\end{table}

\end{widetext}

\section{Conclusions}

We have derived the $\alpha(Z\alpha)^5\,m^2/M$ recoil correction in muonic atoms with
inclusion of the electron vacuum polarization, using the HPQED approach. Splitting the calculation into
low-, middle-, and high-energy regions, all divergent terms, including the double poles
$1/\epsilon^2$ that arise from the vacuum-polarization insertion into the exchanged photons,
cancel in the sum. The result consists of the Salpeter formula evaluated with the
Uehling-modified wave function, Hamiltonian, and Bethe logarithm, and of the additional
contribution $\delta E^{(5)}$ given by extra local terms and new effective potentials, Eq.~(\ref{58}), which we
obtained in closed form.

For the $2P-2S$ Lamb shift the combined electron vacuum-polarization and recoil effect amounts to
$0.18\,\mu$eV in $\mu$H and $0.07\,\mu$eV in $\mu$D, consistent with the previous
logarithmic estimate.
Our results for the muonic helium ions of $-1.26\,\mu$eV in $\mu^3$He$^+$ and $-1.01\,\mu$eV in
$\mu^4$He$^+$ are of the opposite sign and about four times
smaller in magnitude than the logarithmic estimate \cite{JentschuraWundt2011}. Our analytic formula reproduces the
$\ln^2(\beta Z\alpha)$ term of Ref.~\cite{JentschuraWundt2011} but not the single-logarithmic term. Since
$\ln(m/m_e)\approx 5.3$ is not a large parameter and $\beta$ is of order one, the
nonlogarithmic terms and the effective potential $V_\mathrm{SVP}$ are as important as the
logarithms. Logarithmic estimates of combined electron vacuum-polarization and recoil
effects in muonic atoms should therefore be treated with caution.

In all studied systems, the combined effect of evp and nuclear recoil is well below the present experimental uncertainties of
the $2P-2S$ transitions and the uncertainties due to nuclear structure \cite{Pachucki2024,Krauth2021,Schuhmann2025}. Its importance lies in
replacing a logarithmic estimate by a complete result, so that this contribution no longer limits the QED part of the theory. 
Removing theoretical uncertainties from the point-nucleus QED theory
is important for possible improved determinations of nuclear charge radii in the future.
The approach can be applied directly to other muonic ions, such as muonic lithium and beryllium, 
where precision X-ray spectroscopy of the $2P$--$1S$ transitions is currently being pursued \cite{QUARTET}.  
Further improvement of the theory of muonic atoms is limited by the not well known nuclear polarizability correction,
in particular in the three-photon exchange.

\appendix

\section{VP Potentials} \label{appendix:A}

Here, we present the derivation of the vacuum-polarization potentials used
in $E_\mathrm{evp,rec}^{(5)}$. Defining
$\lambda = \beta\,\xi$
and
\begin{align}
	V_{\mathrm{VP}j} =&\
\frac{\alpha}{\pi}\int_2^\infty d\xi\,\frac{2}{3\,\xi}\sqrt{1-\frac{4}{\xi^2}}\bigg(1+\frac{2}{\xi^2}\bigg)
V_j(\lambda)\,,
\end{align}
we get
\begin{align}
V_\mathrm{2}(\lambda) =&\ \int\frac{d^3k}{(2\pi)^3}e^{i\vec k\cdot\vec r}\,\frac{4\pi}{k^2}\,\ln\bigg(\frac{  k^2}{\lambda^2}+1 \bigg)\,,
\\
V'_\mathrm{2}(\lambda) =&\ -\frac{2}{\lambda}\int\frac{d^3k}{(2\pi)^3}e^{i\vec k\cdot\vec r}\,\frac{4\pi}{k^2+\lambda^2}
\nonumber \\ =&\ -\frac{2}{\lambda}\,\frac{e^{-\lambda\,r}}{r}\,,
\\
V_\mathrm{2} =&\ -\int_\lambda^\infty d\lambda\,V'_2(\lambda)
\nonumber \\ =&\
\frac{2}{r}\int_\lambda^\infty \frac{d\lambda}{\lambda}\,e^{-\lambda\,r}
\nonumber \\ =&\ -
\frac{2}{r}\,\mathrm{Ei}(-\lambda\,r)\,.
\end{align}
Further,
\begin{align}
V_\mathrm{3}(\lambda) =&\
\int\frac{d^3k}{(2\pi)^3}e^{i\vec k\cdot\vec r}\;
\frac{4\pi\,\lambda^2}{k^4}\bigg[\ln\bigg(\frac{  k^2}{\lambda^2}+1 \bigg) - \frac{  k^2}{\lambda^2} \bigg]\,,
\\
\bigg(\frac{V_\mathrm{3}}{\lambda^2}\biggr)' =&\ 
\frac{2}{\lambda^3}\,\int\frac{d^3k}{(2\pi)^3}e^{i\vec k\cdot\vec r}\,\frac{4\pi}{k^2+\lambda^2}
\nonumber \\ =&\
\frac{2}{\lambda^3}\,\frac{e^{-\lambda\,r}}{r}\,,
\\
V_\mathrm{3} =&\ -\lambda^2\int_\lambda^\infty d\lambda\,\bigg(\frac{V_3}{\lambda^2}\biggr)' 
\nonumber \\ =&\
-\frac{2\,\lambda^2}{r}\,\int_\lambda^\infty \frac{d\lambda}{\lambda^3}\,e^{-\lambda\,r}
\nonumber \\ =&\
\lambda^2\,r\,\bigg( \mathrm{Ei}(-\lambda\,r) + \frac{e^{-\lambda\,r}}{\lambda\,r} -\frac{e^{-\lambda\,r}}{\lambda^2\,r^2}\bigg) \,.
\end{align}
Finally,
\begin{align}	
V_\mathrm{4}(\lambda) =&\
\int\frac{d^3k}{(2\pi)^3}e^{i\vec k\cdot\vec r}\,\frac{4\pi}{k^2+\lambda^2}
\,\ln\bigg(\frac{  k^2}{\lambda^2}+1 \bigg)\,,	 \\
V_\mathrm{4}(\lambda_1,\lambda_2) =&\
\int\frac{d^3k}{(2\pi)^3}e^{i\vec k\cdot\vec r}\,\frac{4\pi}{k^2+\lambda_1^2}
\,\ln\bigg(\frac{  k^2}{\lambda_2^2}+1 \bigg)\,,	 \\ \nonumber \\	
\frac{d V_\mathrm{4}}{d\,\lambda_2}=&\ 	
-\frac{2}{\lambda_2}\,\int\frac{d^3k}{(2\pi)^3}e^{i\vec k\cdot\vec r}\,\frac{4\pi}{k^2+\lambda_1^2}\,\frac{k^2}{k^2+\lambda_2^2}
\nonumber \\ =&\
-\frac{2}{\lambda_2}\,\int\frac{d^3k}{(2\pi)^3}e^{i\vec k\cdot\vec r}\,\frac{4\pi}{\lambda_1^2-\lambda_2^2}
\,\biggl(\frac{\lambda_1^2}{k^2+\lambda_1^2} \nonumber \\ &\ -\frac{\lambda_2^2}{k^2+\lambda_2^2}\bigg)
\nonumber \\ =&\
-\frac{2}{\lambda_2}\,\frac{1}{\lambda_1^2-\lambda_2^2}\,\bigg(\lambda_1^2\,\frac{e^{-\lambda_1\,r}}{r} - \lambda_2^2\,\frac{e^{-\lambda_2\,r}}{r}\bigg)\,,
\\
V_\mathrm{4} =&\ - \frac{e^{-\lambda\,r}}{r}\,\big[\gamma + e^{2\,\lambda\,r}\,\mathrm{Ei}(-2\,\lambda\,r) + \ln(\lambda\,r/2) \big]\,.
\end{align}	

\section{Fourier transform of $r^{-3}$} \label{appendix:B}

For the Araki--Sucher term, we use the following Fourier transform:
\begin{align}
\int d^3r\,\frac{e^{i\,\vec k\cdot\vec r}}{r^3} \equiv&\ \lim_{\epsilon \rightarrow 0} \bigg[\int_\epsilon d^3r\,\frac{e^{i\,\vec k\cdot\vec r}}{r^3} +4\,\pi\,(\ln\epsilon +\gamma)\bigg]
\nonumber \\ =&\
\lim_{\epsilon \rightarrow 0} \bigg[2\,\pi\int_\epsilon^\infty \frac{dr}{r}\int_{-1}^1 dx\,e^{i\,k\,r\,x} +4\,\pi\,(\ln\epsilon +\gamma)\bigg]
\nonumber \\ =&\
\lim_{\epsilon \rightarrow 0} 4\,\pi\,\bigg[\int_\epsilon^\infty \frac{dr}{r}\,\frac{\sin(k\,r)}{k\,r} +\ln\epsilon +\gamma\bigg]
\nonumber \\ =&\
\lim_{\epsilon \rightarrow 0} 4\,\pi\,\bigg[\int_{\epsilon\,k}^\infty dx\,\frac{\sin x}{x^2} +\ln\epsilon +\gamma\bigg]
\nonumber \\ =&\
\lim_{\epsilon \rightarrow 0} 4\,\pi\,\bigg[1-\gamma-\ln(\epsilon\,k) +\ln\epsilon +\gamma\bigg]
\nonumber \\ =&\
4\,\pi\,(1-\ln k)\,.
\end{align}

\end{document}